\documentclass[a4paper,10pt,twocolumn]{article}

\usepackage[english]{babel}
\usepackage[utf8]{inputenc}
\usepackage[T1]{fontenc}

\usepackage[top=1.5cm, left=1.5cm, right=1.5cm, bottom=1.5cm]{geometry}

\renewenvironment{abstract}{\bf\small {\em\ Abstract---}}{}

\usepackage{amsfonts,amssymb,amsmath,amsthm}
\usepackage{subfigure}
\usepackage{graphicx}
\usepackage[footnotesize]{caption}

\usepackage{amsfonts,amssymb,amsmath,amsthm}

\usepackage[pageanchor=false,colorlinks=true,linkcolor=blue,citecolor=blue,urlcolor=blue]{hyperref}
\usepackage{enumitem}

\newcommand\bR{\mathbb{R}}
\newcommand\bRp{\bR_{+}}

\newcommand\sH{\mathcal{H}}
\newcommand\sA{\mathcal{A}}
\newcommand\sC{\mathcal{C}}

\newcommand\sT{\mathcal{T}}

\newcommand\cl[1]{\overline{#1}}

\newcommand{\tconv}{\mathrm{conv}}

\newcommand{\tspan}{\mathrm{span}}

 \newcommand{\vol}{\mathrm{vol}}

\newcommand{\defin}{:=}
\newtheorem{theorem}{Theorem}[section]

\newtheorem{definition}{Definition}[section]

\title{Is the 1-norm the best convex sparse regularization?\vspace*{-5mm}}

\author{Yann Traonmilin$^1$, Samuel Vaiter$^2$, Rémi Gribonval$^3$.\\
  {\footnotesize $^1$CNRS, Institut de Mathématiques de Bordeaux, Talence, France; $^2$CNRS, Institut de Mathématiques de Bourgogne, Dijon, France;}\\ {\footnotesize $^3$Univ Rennes, Inria, CNRS, IRISA.}
} \date{\empty} 

\begin{document}
\maketitle
\begin{abstract} 
The 1-norm is a good convex regularization for the recovery of sparse vectors from under-determined linear measurements. No other convex regularization seems to surpass its sparse recovery performance. How can this be explained? To answer this question, we define several notions  of ``best'' (convex) regularization in the context of general low-dimensional recovery and show that indeed the 1-norm is an optimal convex sparse regularization within this framework.  
\end{abstract}

\section{Introduction}
We consider the observation model in a Hilbert space $\sH$ (with associated norm $\|\cdot\|_\sH$):
\vspace*{-0.5em}
\begin{equation}
 y = Mx_0
\end{equation}
where $M$ is an under-determined linear operator,  $y$ is a $m$-dimensional vector and $x_0$ is the unknown. We suppose that $x_0$ belongs to a low-dimensional model $\Sigma$ (a union of subspaces). We consider the following minimization program.
\vspace*{-0.5em}

\begin{equation} \label{eq:minimization1}
 x^* \in \arg \min_{Mx=y} R(x)
\end{equation}
where $R$ is a regularization function. A huge body of work gives practical regularizations ensuring that $x^* = x_0$ for several low-dimensional models (in particular sparse and low rank models, see \cite{Foucart_2013} for a most complete review of these results) and convex regularizations. The operator $M$ is generally required to satisfy some property (e.g., the restricted isometry property (RIP)) to guarantee recovery.  In this work, we aim at finding the ``best'' convex regularization for exact recovery of $x_0 \in \Sigma$.  
\vspace*{-1em}
\paragraph{Best regularization with respect to a low dimensional model.}
We describe the framework to define what is the ``best'' regularization in a set of convex functions $\sC$ that was initiated in \cite{Traonmilin_2018} (This work is a follow-up of this article\footnote{The full version of \cite{Traonmilin_2018} with proofs is avalaible at \url{https://hal.inria.fr/hal-01720871}}). If we do not have prior information on $M$, we want to build a compliance measure  $A_{\Sigma}(R)$ that summarizes the notion of good regularization with respect to $\Sigma$ and maximize it
\vspace*{-0.5em}
\begin{equation}
R^* \in \arg\max_{R \in \sC} A_{\Sigma}(R). 
\end{equation}
In the sparse recovery example studied in this article, the existence of a maximum of $A_{\Sigma}(R)$ is verified. However, we could ask ourselves what conditions on $A_\Sigma(R)$ and $\sC$ are necessary and sufficient for the existence of a maximum, which is out of the scope of this article.  
\vspace*{-1em}
\paragraph{Compliance measures.}
When studying recovery with a regularization function $R$, two types of guarantees are generally used: uniform and non-uniform. To describe these recovery guarantees, we use the following definition of descent vectors. 	

 \begin{definition}[Descent vectors]
 For any $x \in \sH$, the collection of descent vectors of $R$ at $x$ is
 \vspace*{-0.5em}
 \begin{equation}
 \sT_{R}(x) \defin \left\{ z \in \sH : R(x+z) \leq R(x) \right\}.
 \end{equation}
\end{definition}
We write $\sT_R(\Sigma):=\bigcup_{x\in \Sigma} \sT_R(x)$. When $R$ is convex these sets are cones. Recovery is characterized by descent vectors (recall that $x^*$ is the result of minimization~\eqref{eq:minimization1}): 
\begin{itemize}[leftmargin=*]
 \item Uniform recovery: Let $M$ be a linear operator. Then  ``for all $x_0 \in \Sigma$, $x^* =x_0$'' is equivalent to $\sT_R(\Sigma) \cap \ker M =\{0\}$.
 \item Non-uniform recovery: Let $M$ be a linear operator and $x_0 \in \Sigma$.  Then $x^* =x_0$  is equivalent to $\sT_R(x_0) \cap \ker M = \{0\}$.
\end{itemize}
Hence, a regularization function $R$ is ``good'' if  $\sT_R(\Sigma)$ leaves a lot of space for $\ker M$ to not intersect it (trivially). In dimension $n$, if there is no orientation prior on the kernel of $M$, the amount of space left can be quantified by the ``volume'' of $\sT_R(\Sigma) \cap S(1)$ where $S(1)$ is the unit sphere with respect to $\|\cdot\|_\sH$. Hence, in dimension $n$, we define a compliance measure for uniform recovery as: 
\vspace*{-1em}
\begin{equation}
A_\Sigma^U(R) := 1 - \frac{\vol\left(\sT_R(\Sigma) \cap S(1)\right)}{\vol(S(1))}.
\end{equation}
More precisely, here, the volume $\vol(E)$ of a set $E$ is the measure of $E$ with respect to the uniform measure on the sphere $S(1)$ (i.e. the $n-1$-dimensional Haussdorf measure of $\sT_R(\Sigma) \cap S(1)$). When looking at non-uniform recovery for random Gaussian measurements, the quantity $\frac{\vol\left(\sT_R(x_0) \cap S(1)\right)}{\vol(S(1))}$ represents the probability that a randomly oriented kernel of dimension 1 intersects (non trivially) $\sT_R(x_0)$. The highest probability of intersection with respect to $x_0$  quantifies the lack of compliance of $R$, hence we can define: 
\vspace*{-1em}
\begin{equation}\label{eq:anusr}
A_\Sigma^{NU}(R) := 1 - \sup_{x \in \Sigma} \frac{\vol\left(\sT_R(x) \cap S(1)\right)}{\vol(S(1))}
\end{equation}
Note that this can be linked with the Gaussian width and statistical dimension theory of sparse recovery \cite{Chandrasekaran_2012, Amelunxen_2014}. 
In infinite dimension, the volume of the sphere $S(1)$ vanishes, making the measures above uninteresting. However, \cite{Traonmilin_2016} and \cite{Puy_2015} show that we can often come back to a low-dimensional recovery problem in an intermediate finite (potentially high dimensional) subspace of $\sH$. Adapting the definition of $S(1)$ to this subspace allows to extend these compliance measures. 

While it was shown that the $\ell^1$-norm is indeed the best atomic norm for $A_\Sigma^U(R)$ and $A_\Sigma^{NU}(R)$ in the minimal case of 1-sparse recovery for $n=3$ in \cite{Traonmilin_2018}, extending these exact calculations to the case of $k$-sparse recovery in dimension $n$ seems out of reach.
\vspace*{-1em}
\paragraph{Compliance measures based on the RIP.}\label{sec:RIP_compliance}

For uniform recovery, another possibility is to use recovery results based on the restricted isometry property. They have been shown to be adequate for multiple models~\cite{Traonmilin_2016}, to be tight in some sense for sparse and low rank recovery~\cite{Davies_2009}, to be necessary in some sense~\cite{Bourrier_2014} and to be well adapted to the study of random operators \cite{Puy_2015}. 

\begin{definition}[RIP constant] 
 Let $\Sigma$ be a union of subspaces and $M$ be a linear map, the RIP constant of $M$ is defined as 
 \vspace*{-0.5em}
 \begin{equation}
  \delta(M) = \sup_{x \in \Sigma-\Sigma} \left| \frac{\|Mx\|_\sH^2}{\|x\|_\sH^2}-1 \right|,
 \end{equation}
 where $\Sigma-\Sigma$ (differences of elements of $\Sigma$) is called the secant set. 
 \end{definition}
 It has been shown that if $M$ has a RIP with constant $\delta< \delta_\Sigma(R)$ on the secant set $\Sigma-\Sigma$, with $\delta^{\mathtt{suff}}_\Sigma(R)$ being fully determined by $\Sigma$ and $R$~\cite{Traonmilin_2016}, then uniform stable recovery is possible. The explicit constant  $\delta_\Sigma^{\mathtt{suff}}(R)$ is only sufficient (and sharp in some sense for sparse and low rank recovery). An ideal RIP based compliance measure would be to use a sharp RIP constant $\delta^{\mathtt{sharp}}_\Sigma(R)$ (unfortunately, it is an open question to derive analytical expressions of this constant for sparsity and other low-dimensional models) defined as: 
\begin{equation}
 \delta^{\mathtt{sharp}}_\Sigma(R) := \inf_{M : \ker M \cap \sT_R(\Sigma) \neq \{0\} } \delta(M).
\end{equation}
It is the best RIP constant of measurement operators where uniform recovery fail. When $ \delta^{\mathtt{sharp}}_\Sigma(R) $ increases, $R$ permits recovery of $\Sigma$ for more measurement operators $M$ (less stringent RIP condition). Hence $ \delta^{\mathtt{sharp}}_\Sigma(R) $ can be viewed as a compliance measure:
\begin{equation}
 A^{RIP}_\Sigma(R)= \delta^{\mathtt{sharp}}_\Sigma(R).
\end{equation}
The lack of practical analytic expressions for $\delta^{\mathtt{sharp}}_\Sigma(R)$ limits the possibilities of exact optimization with respect to $R$.
We propose to look at two RIP based compliance measures:  
\begin{itemize}[leftmargin=*]
 \item  A measure based on necessary RIP conditions \cite{Davies_2009} which yields sharp recovery constants for particular operators, e.g.,
\begin{equation}
 A_\Sigma^{RIP,\mathtt{nec}}(R)=\delta^{\mathtt{nec}}_\Sigma(R) := \inf_{z \in \sT_R(\Sigma) \setminus \{0\}} \delta(I-\Pi_z ).
\end{equation}
 where $\Pi_{z}$ is the orthogonal projection onto the one-dimensional subspace $\tspan(z)$  (other intermediate necessary RIP constants can be defined). Another open question is to determine whether  $\delta_\Sigma^{\mathtt{nec}}(R) =  \delta^{\mathtt{sharp}}_\Sigma(R)$ generally or for some particular models.
 \item A measure based on sufficient RIP constants for recovery, i.e. $A_\Sigma(R)^{RIP,\mathtt{suff}}=\delta_\Sigma^{\mathtt{suff}}(R)$ from \cite{Traonmilin_2016}.
\end{itemize}
Note that we have the relation 
 \begin{equation}
 \delta_\Sigma^{\mathtt{suff}}(R) \leq  \delta^{\mathtt{sharp}}_\Sigma(R)  \leq \delta_\Sigma^{\mathtt{nec}}(R). 
\end{equation}
To summarize, instead of considering the most natural RIP-based compliance measure (based on $\delta^{\mathtt{sharp}}_\Sigma(R)$ ), we use the best known bounds of this measure. Moreover, in \cite[Lemma~2.1]{Traonmilin_2016}, it has been shown that given a coercive convex regularization $R$, there is always a  atomic norm $\|\cdot\|_\sA$ (always convex) with atoms $\sA$ included in the model such that $\sT_{\|\cdot\|_\sA}(\Sigma) \subset \sT_{R}(\Sigma)$. 

\begin{definition}
 The atomic ``norm'' induced by the set $\sA$ is defined as: 
\begin{equation}
 \|x\|_\sA \defin \inf \left\{ t \in \bRp:  x \in t\cdot\cl{ \tconv}(\sA) \right\}
\end{equation}
where $\cl{ \tconv}(\sA)$ is the closure of the convex hull of $\sA$. 
\end{definition}

This implies that $A_\Sigma^{U}(\|\cdot\|_\sA) \geq A_\Sigma^{U}(R)$. In consequence, we look for best regularisations in the set $\sC_\Sigma := \{ R: R(x) = \|x\|_\sA, \sA \subset \Sigma, \max_{a\in \sA} \|a\|_2 =1\}$.

\section{Optimality of the $\ell^1$-norm for RIP-based compliance measures}

\label{sec:nec_RIP}

We set $\Sigma = \Sigma_k$  and $\sH = \bR^n$ with $k\geq1$ and  $n\geq 3$. Hence $\Sigma - \Sigma = \Sigma_{2k}$.
 It is possible to show~\cite{Traonmilin_2018}:  
\begin{equation}
\begin{split}
 \arg \max_{R \in \sC_\Sigma} A_\Sigma^{RIP,\mathtt{nec}}(R)  =  \arg \min_{R \in \sC_\Sigma} B_\Sigma(R)  \\ 
 \end{split}
\end{equation}
where   $B_\Sigma(R) := \underset{z\in \sT_R(\Sigma) \setminus \{0\} }{\sup}  \frac{\|z_{T_2^c}\|_2^2}{\|z_{T_2}\|_2^2} $ and $T_2$ is a notation for the support of $2k$ biggest coordinates in $z$, i.e. for all $i \in T_2, j\in T_2^c$, we have $|z_i|\geq |z_j|$. 

Similarly to the necessary case,  we can show 
\begin{equation}
\begin{split}
 \arg \max_{R \in \sC_\Sigma} A_\Sigma^{RIP,\mathtt{suff}}(R)  =  \arg \min_{R \in \sC_\Sigma} D_\Sigma(R)  \\ 
 \end{split}
\end{equation}
where $D_\Sigma(R) := \underset{z\in \sT_R(\Sigma) \setminus \{0\} }{\sup}  \frac{\|z_{T^c}\|_\Sigma^2}{\|z_T\|_2^2} $ and $T$ denotes  the  support of the  $k$ biggest coordinates of $z$. The norm $\|\cdot\|_\Sigma$ is the atomic norm generated by the set of atoms $\Sigma\cap S(1)$. Remark the similarity between the fundamental quantity to optimize for the necessary case and the sufficient case, $B_\Sigma(R)$ and $D_\Sigma(R)$, this leads us to think that our control of $A_\Sigma^{RIP}(R)$ is rather tight. Optimizing $B_\Sigma(R) $ and $D_\Sigma(R) $ for $R \in \sC_\Sigma$ gives the result:

\begin{theorem} \label{th:opt_l1_RIP_atom}
  Let  $n\geq 2k$, $\Sigma = \Sigma_k$, $\sH = \bR^n$ and $\sC_\Sigma = \{ R: R(x) = \|x\|_\sA, \sA \subset \Sigma, \max_{a\in \sA} \|a\|_2 =1\}$. We have
 \begin{equation}
 \begin{split}  
 \|\cdot\|_1 &\in \arg \max_{R\in \sC_\Sigma} A_\Sigma^{RIP,\mathtt{nec}}(R).\\
 \|\cdot\|_1 &\in \arg \max_{R\in \sC_\Sigma} A_\Sigma^{RIP,\mathtt{suff}}(R).\\
 \end{split}
 \end{equation}
\end{theorem}
Note that contrary to \cite{Traonmilin_2018} where multiples of the $\ell^1$-norm  where the sole maximizers of these compliance measures among weighted $\ell^1$-norm, unicity among atomic norms has yet to be proven. 

\section{Discussion and future work}
We have shown that, not surprisingly, the $\ell^1$-norm is an optimal convex regularization for sparse recovery within this framework. The important point is that we could explicitly quantify  a notion of good regularization. This is promising for the search of optimal regularizations for more complicated low-dimensional models such as ``sparse and low rank'' models or hierarchical sparse models.  We also expect similar results for low-rank recovery and the nuclear norm as technical tools are very similar. 

We used compliance measures based on (uniform) RIP recovery guarantees to give results for the general sparse recovery case, it would be interesting to do such analysis using  (non-uniform) recovery guarantees based on the statistical dimension or Gaussian width of the descent cones \cite{Chandrasekaran_2012,Amelunxen_2014}. 

Finally, while these compliance measures are designed to make sense with respect to known results in the area of sparse recovery, one might design other compliance measures tailored for particular needs (e.g. structured operators $M$), in this search for optimal regularizations.
\section*{Acknowledgements}
This work was partly supported by the CNRS PEPS JC 2018 (project on efficient regularizations). 

\bibliographystyle{abbrv}
\bibliography{optimal_reg_itwist2018}

\end{document}